\documentstyle[11pt,epsf]{article}

\newcommand{\bd}{\begin{description}}
\newcommand{\ed}{\end{description}}
\newcommand{\ben}{\begin{enumerate}}
\newcommand{\een}{\end{enumerate}}
\newcommand{\be}{\begin{equation}}
\newcommand{\ee}{\end{equation}}
\newcommand{\bi}{\begin{itemize}}
\newcommand{\ei}{\end{itemize}}

\newcommand{\bt}{\begin{tabular}}
\newcommand{\et}{\end{tabular}}
\newcommand{\bc}{\begin{center}}
\newcommand{\ec}{\end{center}}

\newcommand{\beq}{\begin{eqnarray}}
\newcommand{\eeq}{\end{eqnarray}}
\newcommand{\nn}{\nonumber}

\begin{document}
\begin{titlepage}
\begin{flushright}
 KUNS-1994 \\

\end{flushright}

\bc
\vspace*{10mm}

{\LARGE \bf{
KKLT type models with moduli-mixing superpotential
}}

\vspace{12mm}

{\large
Hiroyuki~Abe\footnote{E-mail address:
  abe@gauge.scphys.kyoto-u.ac.jp}
,~Tetsutaro~Higaki\footnote{E-mail address:
  tetsu@gauge.scphys.kyoto-u.ac.jp}~and~
Tatsuo~Kobayashi\footnote{E-mail address: kobayash@gauge.scphys.kyoto-u.ac.jp}
}

\vspace{6mm}

{\it Department of Physics, Kyoto University,
Kyoto 606-8502, Japan}

\vspace*{15mm}

\begin{abstract}

We study KKLT type models with moduli-mixing superpotential.
In several string models, gauge kinetic functions are 
written as linear combinations of two or more moduli fields.  
Their gluino condensation generates moduli-mixing superpotential.
We assume one of moduli fields is frozen already around the string scale.
It is found that K\"ahler modulus can be stabilized 
at a realistic value
without tuning
3-form fluxes
because of gluino condensation on (non-)magnetized D-brane.
Furthermore, 
we do not need to highly 
tune parameters in order to realize a weak gauge coupling and
a large hierarchy between the gravitino mass and the Planck scale,
when there exists non-perturbative effects on D3-brane.  
SUSY breaking patterns in our models have a rich structure.
Also, some of our models have cosmologically important 
implications, e.g., on the overshooting problem 
and the destabilization problem due to finite temperature effects
as well as the gravitino problem and the moduli problem.
\end{abstract}

\ec

\end{titlepage}

\newpage

\section{Introduction}


String/M theory is a promising candidate for 
unified theory including gravity.
Its 4D  effective theory, in general, 
includes many moduli fields.
Their vacuum expectation values (VEVs) play an important role 
in particle physics and cosmology.
That is because  those VEVs determine coupling constants 
such as gauge and Yukawa couplings, and physical scales 
like the Planck scale $M_p$ and the compactification scale.
How to stabilize moduli (at realistic values) is one of 
important issues to study in string phenomenology 
and cosmology.
How to break supersymmetry (SUSY) is another important issue.
It is expected that nonperturbative effects, which 
fix moduli VEVs, may also break SUSY.
A number of studies on these issues have been done, 
and some of those vacua correspond to anti-de Sitter (AdS) 
vacua.

Recently, flux compactification is studied intensively, 
because it can stabilize some of moduli.
In type IIB string models, 
complex structure moduli and the dilaton can be stabilized 
around the string scale $M_{string}=\alpha '^{-1/2}$, which is
expected to be of ${\mathcal O}(10^{17}) - {\mathcal O}(10^{18})$ GeV, but 
K\"ahler moduli fields are not stabilized \cite{Giddings:2001yu}.
On the other hand, in type IIA string models 
all of moduli can be stabilized \cite{DeWolfe:2005uu}, and in 
heterotic string models complex structure and 
volume moduli can be stabilized, but the dilaton 
VEV is not fixed.
However, in the heterotic case, the compact space is 
generically non-K\"ahler, and it is 
mathematically difficult to treat it by present knowledge.\footnote{
Recently, a solution that compact space is conformally 
K\"ahler is found in Ref.~\cite{Becker:2005nb}.}

Within the framework
of flux compactification in type IIB string models, 
a simple model has been proposed 
in Ref.~\cite{Kachru:2003aw}, 
where 
the remaining K\"ahler modulus $T$ 
is stabilized by nonperturbative 
effects such as gluino condensation.
Still, the potential minimum corresponds to the SUSY AdS vacuum 
at the first stage.
Then, an anti-D3 ($\overline{ \rm D3}$) brane is introduced (at the tip of 
throat) in order to uplift the vacuum energy and 
realize a de Sitter (dS) or Minkowski vacuum.
It shifts the potential minimum, and breaks 
SUSY in a controllable way.
That is the KKLT scenario.

Furthermore, soft SUSY breaking terms in 
the KKLT scenario have been 
studied in Ref.~\cite{Choi:2004sx,Choi:2005ge}, 
and it has been found that 
in the KKLT scenario 
the $F$-term of modulus field $T$ is of 
${\mathcal O}(m_{3/2}M_p /4\pi^2)$, 
where $m_{3/2}$ is the gravitino mass.
That is, the  modulus $F$-term $F^T$ and 
the anomaly mediation \cite{Randall:1998uk,Giudice:1998xp} 
are comparable in 
soft SUSY breaking terms.
That leads to a quite novel pattern of SUSY breaking 
terms.(See for their phenomenological aspects 
Ref.~\cite{Choi:2005uz,Endo:2005uy,Falkowski:2005ck}.)
It is useful to introduce the parameter $\alpha$ as 
$\alpha \equiv \frac{m_{3/2}(T + \overline T)}{F^T 
\ln (M_p/m_{3/2} )}$ \cite{Choi:2005uz}, in order to 
represent the ratio between the modulus mediation and 
the anomaly mediation.
One of phenomenologically interesting features 
in the modulus-anomaly mixed mediation is 
the appearance of a mirage messenger scale \cite{Choi:2005uz}, 
where the anomaly mediation at the GUT scale can 
cancel the renormalization group effect under a certain condition.
The mirage messenger scale $\Lambda_m$ is estimated as 
$\Lambda_m \sim (m_{3/2}/M_p)^{\alpha /2}M_{GUT}$.
That is, soft SUSY breaking terms appear as 
the pure modulus mediation at the mirage scale.
The mirage scale depends on the ratio between 
the modulus $F$-term and $m_{3/2}$.
The original KKLT model leads to $\alpha \approx 1$, 
and  the mirage scale is the intermediate scale.

Moreover, it has been pointed out in \cite{Choi:2005hd} 
that if 
the mirage scale is around ${\mathcal O}(1)$ TeV, 
such model has an important implication 
on solving the little SUSY hierarchy problem.
At any rate, the mirage scale is determined by 
the ratio $\alpha$, 
which has phenomenologically interesting 
implications.

The KKLT model is also interesting in cosmology.
The anomaly mediation is sizable, 
that is, the gravitino mass is ${\mathcal O}(10)$ TeV.
The modulus is much heavier.
Thus, we may avoid the cosmological gravitino/moduli
problem \cite{Choi:2005ge,Choi:2005uz,Endo:2005uy,Falkowski:2005ck}.

In this paper, we study a modified KKLT scenario.
A gauge kinetic function is a mixture of two 
or more moduli fields in several string models, 
e.g., weakly coupled heterotic string models 
\cite{Choi:1985bz,Ibanez:1986xy}, 
heterotic M models \cite
{Banks:1996ss,Choi:1997an,Nilles:1997vk,Buchbinder:2003pi,Curio:2000dw}, 
type IIA intersecting 
D-brane models and type IIB magnetized D-brane 
models \cite{Cremades:2002te,Lust:2004cx}.
Suppose that the gauge kinetic function $f$ is 
a linear combination of the dilaton $S$ 
and the modulus $T$ like $f = m S + wT$.
Following Ref.~\cite{Kachru:2003aw}, 
we assume that $S$ is frozen already around $M_{string}$ \footnote{
Alternatively, both of them may remain light in some models.
We will study such models separately in Ref.~\cite{AHK}.}.
We also assume 
gluino condensation generates 
nontrivial $T$-dependent superpotential, which
form is expected to be $e^{-cf} \sim Be^{-bT}$.
The original KKLT model corresponds to the case that 
$B={\mathcal O}(1)$ and $b$ is positive.
However, in our case the constant 
$B$ can be very suppressed depending on 
$mc\langle S \rangle $, and also 
the exponent coefficient $b$ can be negative even though 
it is generated by asymptotically free 
gauge sector \footnote{
A similar superpotential like $e^{bT}$ with $b>0$ 
has been studied for non-asymptotically free models 
in Ref.~\cite{Burgess:1997pj}.
However, we stress that 
our superpotential is generated by 
asymptotically free gauge sector.
}.
We study moduli stabilization and 
SUSY breaking with such potential terms, by adding 
uplifting potential.
One feature of such SUSY breaking is that 
the ratio between $F^T/(T + \bar T)$ and $m_{3/2}$
can vary by a value of $B$.
That is phenomenologically interesting,
because we would have richer structure of SUSY spectra.
Moreover, the superpotential term with $b< 0$ has 
important implications on cosmology.
It may avoid the overshooting problem and 
the destabilization problem due to finite temperature effects.

This paper is organized as follows.
In section 2, we review on flux compactification 
and the KKLT model.
In section 3, we give concrete string/M models, where 
gauge kinetic functions depend on two or more moduli fields.
Then, in section 4 we study moduli stabilization 
and SUSY breaking.
In section 5, we discuss implications on 
SUSY phenomenology and cosmology.
Section 6 is devoted to conclusion.
In Appendix A, detailed analysis on the potential 
minimum is summarized.

\section{
Review on flux compactification and KKLT model
}

\subsection{Flux compactification}

In this subsection, we review on flux compactification 
and in the next subsection we review on the 
KKLT model.

We consider the Type IIB $O3/O7$ 
orientifold 4D string model on  
a warped Calabi-Yau (CY) threefold with $h_{1,1}({\rm CY})=1$. 
In addition, 
we introduce RR and NSNS 3-form fluxes
$F_3^{RR}$ and $H_{3}^{NS}$, which should be quantized 
on compact 3-cycles $C_3$ and $C_3'$ such as
\beq
\frac{1}{2\pi \alpha'}\int_{C_3}F_3^{RR} \in 2\pi {\mathbf Z}, 
\qquad
\frac{1}{2\pi\alpha'}\int_{C_3'}H_3^{NS} \in  2\pi {\mathbf Z}.
\eeq  
Furthermore, these fluxes should satisfy the RR tadpole condition
\beq
\frac{1}{(2\pi)^4\alpha'^2}\int_{M_6}H_3^{NS} \wedge F^{RR}_3
+Q^{local}=0,
\label{rrtad}
\eeq
where $Q^{local}$ is the RR charge contribution of local objects
including D3-brane, wrapped D7-brane and O3-planes 
in the D3-brane charge unit.
In this flux compactified type IIB string model, 
we can fix dilaton and complex structure moduli including 
a warp factor but not the K\"ahler modulus around the 
string scale.
Thus,  only the K\"ahler modulus remains light.

We are interested in two moduli, that is, 
one is the dilaton $S$ and another is the overall K\"ahler modulus 
$T$, although the dilaton is frozen around $M_{string}$.
They are given by
\beq
2\pi S= {e^{-\phi}} -ic_0 , \, \qquad 
 2\pi T= {v^{2/3}_E} - i c_4.
\eeq
Here, $\phi$ is 10D dilaton and $v_E$ is a volume of CY in the string
unit within the Einstein frame.
The 10D Einstein metric is given by the string metric
 $g_{MN}^{string}=e^{\phi/2}g_{MN}^E$.
Thus, the CY volume in the string frame $v$ is written by 
$v=e^{3\phi/2}v_E$.
The axionic mode $c_0$ is the RR scalar and  
$c_4$ is the 4D Poincare dual of (1,1) part of 4-form RR potential. 
For example, these fields are related to gauge kinetic functions on 
D3 and non-magnetized D7 brane $f_{Dp}$
\beq
f_{D3}={S}, \qquad f_{D7}={T},
\eeq
where $\langle Ref_{Dp} \rangle =g^{-2}_{Dp}$ and 
$g_{Dp}$ is the gauge coupling on the Dp-brane. 
In this perturbative description,
it is natural that ${\langle ReS \rangle},~ {\langle ReT \rangle}
={\mathcal O}(1)$. 
In addition, these VEVs are related to the physical
scales
such as the 4D Planck scale $M_p$ and 
the compactification scale $M_{KK}$ \cite{Choi:1997an,Choi:2005ge}
\beq
\frac{M_p}{M_{string}}=2{\sqrt{\pi}}(\langle ReS \rangle
\cdot \langle ReT^3 \rangle)^{1/4},
\qquad
\frac{M_{string}}{M_{KK}}=\left(\frac{\langle ReT \rangle}{\langle ReS
    \rangle}\right)^{1/4},
\label{scales}
\eeq
where $M_{KK}/M_{string} \equiv  \langle v^{-1/6} \rangle$.
We need the condition $\langle ReT \rangle > \langle ReS \rangle$. 

With the 3-form flux and orientifold planes in the compact space,
the dilaton $S$ and the complex structure
moduli are frozen around $M_{string}$
and a
background metric is warped , then the metric in the 10D Einstein frame 
is given by \cite{Giddings:2001yu}
\beq
ds^{2}_{10}=\frac{1}{g_s^2 v_E}e^{2A(y)}
g_{\mu\nu}^{E}dx^{\mu}dx^{\nu}+e^{-2A(y)} v_E^{1/3}
\tilde{g}_{mn}dy^mdy^n.
\label{metric}
\eeq 
Here, 
$g_s = e^{\langle \phi \rangle}$ is string coupling,
$g_{\mu\nu}^{E}$ is 4D Einstein metric and $\tilde{g}_{mn}$ is 
unwarped compact 
CY metric which is 
normalized as $\int d^6y \sqrt{{\rm det}\tilde{g}_{mn}}=
(2\pi\alpha'^{1/2})^6$. 
The $y^m$ dependence of the warp factor $e^{2A(y)}$ 
on the throat  is studied in
\cite{Kachru:2003sx,Klebanov:2000hb} and
in generic point we have $e^{2A(y)}\sim 1$. 
A minimum of warp factor can be treated as complex
structure deformation from CY conifold,
and the warp factor on the tip of the throat
$e^{A_{min}}$ can be stabilized by 3-form flux
such that \cite{Giddings:2001yu,Kachru:2003sx},
\beq
a_0
\equiv 
\exp[(-2\pi h)/(3g_s f)]=  
\frac{e^{A_{min}}}{v_E^{1/6}}. 
\label{warp}
\eeq
Here, $f,h$ are given by RR and NSNS 3-form fluxes
\beq
2\pi f= \frac{1}{2\pi \alpha'}\int_{A}F_3^{RR}, \qquad
-2\pi h= \frac{1}{2\pi \alpha'}\int_{B}H_3^{NS},\qquad f,h \in \mathbf{Z},
\eeq
where $A$ is 3-cycle and $B$ is dual cycle of $A$ near conifold singularity.
Thus, if $h \gg g_s f$, 
we can produce exponentially large hierarchy in this string model,
like the Randall-Sundrum model \cite{Randall:1999ee}.

We consider moduli stabilization within the framework of 
4D $N=1$ effective supergravity.
Here and hereafter,  we use the unit that
  $M_{p}=2.4 \times 10^{18} ~{\rm GeV}=1$.
Here we neglect the warp factor dependence of potential, because
warping effects are not important in generic point 
of the compact CY space
except for
the small region on the warped throat. 
Hence, since moduli are bulk fields, they may not be affected by
warping. 
The stabilization of $S$ and the complex structure moduli 
$U^{\alpha}$ is as follows.
The 
3-form flux in the compact space $M_6$ 
generates the following superpotential,
\cite{Gukov:1999ya}
\beq
W_{flux} &=&W(S,U^{\alpha})=\int_{M_6} G_3 \, \wedge \, \Omega ,
\eeq
where $G_3=F_3^{RR}-2\pi i SH_3^{NS}$ and  $\Omega$ is 
the holomorphic 3-form on
CY.  
With the following K\"ahler potential,
\beq
K&=& -\ln(S+\overline{S}) 
-3\ln(T+\overline{T})-\ln(-i\int_{M_6} \Omega \wedge
\overline{\Omega}),
\eeq
the scalar potential is written as 
\beq
V&=&e^{K}\left(
D_{a}W\overline{D_{b}W}K^{a\overline{b}}-3|W|^2
\right), \nn \\ 
&=&e^{K}\left(
D_{i}W\overline{D_{j}W}K^{i\overline{j}}
\right),
\eeq
where $D_aW=(\partial_a K) W+ \partial_a W$,
$K_{a\overline{b}}=\partial_{a}\partial_{\overline{b}}K$,
$a,b$ are summed over all moduli fields and $i,j$ are moduli
fields excluding $T$ because of no-scale structure. 
The potential minimum, i.e. the $F$-flatness condition, 
is obtained as 
\cite{Giddings:2001yu,Kachru:2002he}
\beq
D_SW&=& \frac{-1}{(S+\overline{S})} \int_{M_6} \overline{G_3} \, \wedge \,
\Omega =0 ,
\label{susy1}
\\
D_{U^{\alpha}}W&=& \int _{M_6} G_3 \, \wedge \, \chi_{\alpha}=0.
\label{susy2}
\eeq
Here, we have used $\partial_{U^{\alpha}}\Omega =
-(\partial_{U^{\alpha}}K)\, 
\Omega + \chi_{\alpha}$, where 
$\chi_{\alpha}$ is a basis of primitive (2,1) forms.\footnote{
The primitivity means
$g^{j\overline{k}}\chi_{ij\overline{k}}=0$.}
Thus, these moduli are generically
stabilized at values of order unity around $M_{string}$, 
but the K\"ahler modulus cannot be
stabilized at this stage. 
The stabilization of the dilaton and the complex structure moduli 
is possible, because the degree of freedom of $G_3$ is equal 
to the number of the dilaton $S$
and complex structure moduli, that is $2(1+h_{2,1})$. 
{}From (\ref{susy1}) and (\ref{susy2}), the 3-form flux $G_3$ 
can be decomposed
to $(0,3)$ form $G_{(0,3)}$ and primitive $(2,1)$ forms $G_{(2,1)}^P$.
As a result, $G_3$ is imaginary self-dual (ISD), such that $*_6
G_{3}=iG_3$. 
That leads to  
\beq
\nn
(2\pi)^4\alpha'^2 N_{\rm flux} &\equiv & \int_{M_6} H^{NS}_3
\wedge F_3^{RR} = \frac{ie^{\phi}}{2}\int_{M_6}  G_3 \wedge \overline{G_3},
\\
&= & \frac{e^{\phi}}{2\cdot 3!}\int_{M_6}d^6y \sqrt{g_6}  \,
G_{mnl}\overline{G_3}^{mnl}
\geq 0,
\eeq 
then from (\ref{rrtad}) we can see that we actually need negative charge
contributions, such as O3-plane, wrapped D7-brane or 
$\overline{{\rm D3}}$-brane etc.
These ISD fluxes stabilize the position of D7-branes and
$\overline{{\rm D3}}$-branes \cite{Camara:2003ku}, when 
the back reaction from $\overline{{\rm D3}}$-branes 
to the background is neglected \cite{Kachru:2002gs}.
The other open string
moduli of D7-branes 
cannot be stabilized in  the ISD flux background. 
SUSY breaking effects (including imaginary anti self-dual fluxes) or
other bulk geometry may stabilize them. 

\subsection{KKLT model}
\label{KKLT}
In this subsection, we review on the KKLT model.
See Appendix A for details of potential analysis.
We assume that the dilaton and complex structure moduli are 
frozen around $M_{string}$ through flux compactification 
as mentioned in the previous subsection.
Here, for simplicity we neglect constant K\"ahler potential of frozen 
dilaton and complex structure moduli because those only change overall
magnitude of scalar potential.
To stabilize the remaining K\"ahler modulus, 
the $T$-dependent superpotential is added in Ref.~\cite{Kachru:2003aw}. 
Such superpotential can be generated by non-perturbative effects 
at a low energy scale in the ISD flux background with $G_{(0,3)} \neq 0$. 
Then  K\"ahler potential and superpotential are written as 
\beq
K&=&-3 \ln (T+\overline{T}),
\label{kpt1}
\\
W&=&w_0 - Ce^{-aT}, 
\label{spt1}
\eeq
where $a=8\pi^2/N$ with $N \in \mathbf{N}$ and ,
\beq
w_0&=&\langle \,W_{flux} \, \rangle =
\langle \, \int_{M}G_{(0,3)} \wedge \Omega \, \rangle . 
\eeq
The second term in R.H.S. of (\ref{spt1}) originates from $SU(N)$ ($N > 1$)
gauge group gluino
condensation on non-magnetized D7-branes wrapping  unwarped 4-cycle
or Euclidean
D3-brane instanton ($N=1$) wrapping a similar cycle.
The constant $C$ can depend on complex structure moduli, 
i.e. $C=C(\langle U^{\alpha} \rangle)$.
For gluino condensation, we have $C = N M_{string}^3$ and for D3-brane
instanton $C$ is 1-loop determinant of D3-brane mode which depend on
complex structure moduli.
Thus, the natural order of $C$ is of ${\mathcal O}(1)$ or suppressed 
by one-loop factor.
This superpotential generates the following scalar potential,
\beq
V_F&=& e^{K}\left[
|D_{T}W|^2\frac{(T+\overline{T})^2}{3}-3|W|^2
\right],
\eeq
and
stabilizes the K\"ahler modulus, such
that $\partial_{T}V_F=D_{T}W=0$, i.e.,
\beq
D_{T}W&=&\frac{-3}{(T+\overline{T})}
\left[
w_{0}-\left( 
1+\frac{a(T+\overline{T})}{3}
\right)Ce^{-aT}
\right]
=0. 
\label{susy3}
\eeq
That corresponds to the  SUSY AdS vacuum 
\beq
\langle V_F \rangle =-3\langle \,  e^{K}|W|^2 \, \rangle.
\eeq
Then, we require $|w_0| \ll  |C|$ and
the VEV of $T$ is obtained as 
\beq
a ReT &=&  \ln \left[
\frac{|C| (3+a ( T  +  \overline{T} ) ) }{3|w_{0}|}
\right]
\simeq
|\ln (w_{0})\,  | \simeq  \ln \left[ \frac{M_p}{m_{3/2}} \right],
\\
a \, ImT&=&-Arg(w_0)+Arg(C)+2\pi n ,~~n\in~ {\mathbf Z},
\eeq 
that is, $a ReT  ={\mathcal O}(4\pi^2)$.
The gravitino mass $m_{3/2}$ and the modulus mass $m_T$ are 
obtained as 
\beq
m_{3/2}^2 \equiv \langle e^K |W|^2 \rangle
=e^{K}\frac{a^2(T+\overline{T})^2}{9}|C|^2e^{-2aT} \simeq
{|w_0|^2} \ll 1,
\label{kklt-gravitino}
\eeq
\beq
m_{T} \simeq a(T+\overline{T})m_{3/2}.
\label{kklt-moduli}
\eeq
In order to obtain suppressed gravitino mass, we need a very small value
of $w_0$, that is, $\langle |W_{flux}|\rangle \ll 1$. 
Study on landscape of flux vacua suggests that 
the number of the vacua $N_{vac}$ with gravitino mass $m_{3/2}$
is typically given by $N_{vac} \sim
m_{3/2}^2\cdot 10^{300}$ \cite{Denef:2004ze, Gorlich:2004qm}. 
This number is amazingly large, 
and we can tune fluxes such that $|w_0| \ll 1$. 
For example \cite{Choi:2005ge}, 
it is tuned as $|G_{(0,3)}/G_{(2,1)}| \ll 1$. 

To realize a dS or Minkowski vacuum, 
we need to uplift the potential by $3m_{3/2}^2 \sim |w_0|^2 \ll 1$.
That has been done in Ref.~\cite{Kachru:2003aw} by 
adding a single $\overline{{\rm D3}}$-brane stabilized on the tip
of warped throat as an origin of uplifting scalar potential.
The following potential 
\beq
V_{L}= 2\frac{a_0^4T_{3}}{g_s^4}\frac{1}{4\pi^2(ReT)^2}
\equiv \frac{D}{(T+\overline{T})^2}, 
\label{uplift}
\eeq
where $T_3=(2\pi)^{-3}(\alpha')^{-2}$ is D3-brane tension, 
is generated from $\overline{{\rm D3}}$ tension
and Wess-Zumino term.
Because of the warp factor $a_0$, 
the uplifting potential $V_{L}$ can be very small.
Then the total scalar potential is written as 
\beq
V= V_F +V_L.
\eeq
In order to have  $\langle V \rangle =0$, 
we have to tune the warp factor using fluxes, 
such as $a_0 = \exp[-2\pi h/(3g_s f)] \sim \sqrt{ m_{3/2} }$ .

The uplifting potential shifts the minimum.
The VEV of $T$ slightly changes and non-vanishing 
$F^T$ is generated.
{}From the above superpotential,  we can estimate $W_{TT} = -aW_T$.
Then, by use of analysis in Appendix A, we evaluate $F^T$ as 
\beq
\left| \frac{F^T}{(T+\overline{T})}\right| &\simeq &\frac{m_{3/2}}{aReT},\\
Arg(\langle F^T \rangle )&=&Arg(\langle \overline{W} \rangle )=-Arg(w_0).
\eeq
It is useful to use the following parameter $\alpha$,
\beq
\alpha &\equiv& \frac{m_{3/2}(T+\overline{T})}{ F^T \ln(M_p/m_{3/2})},
\eeq
in order to represent the ratio between anomaly
mediation and modulus mediation.
In this case we have $\alpha \simeq 1$. 
We can delete phase of superpotential
due to $U(1)_R$ symmetry
$W \rightarrow e^{-iArg({w_0})}W$ and
PQ symmetry $aT \rightarrow aT+i(\, Arg(C)-Arg({w_0}) \, )$
\cite{Choi:2005ge,Choi:1993yd}. 
The results of (\ref{kklt-gravitino}) and (\ref{kklt-moduli}) do not change.

\begin{figure}
\epsfxsize=0.7\textwidth
\centerline{\epsfbox{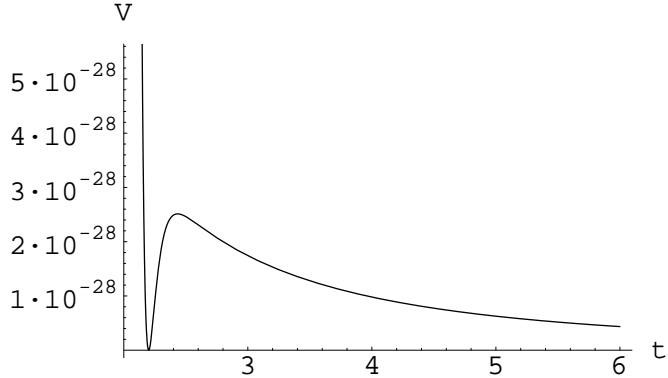}}
\caption{A potential plot of KKLT model
with parameters, $w_0=10^{-13}$, 
$C=N,~a=8\pi^2/N$, 
$N=5$ and $D=6.3\times 10^{-27}$. 
Then, we obtain $\langle ReT \rangle \simeq 2.2$ and $m_{3/2} \simeq 25$ TeV.
A horizontal axis is $t=ReT$ and a vertical axis is
$V$. }
\label{fig1}
\end{figure}

We show an illustrating example in Figure \ref{fig1}.
We take parameters as 
$C=N$, $a=8\pi^2/N$, 
$N=5$ and $D=6.3\times 10^{-27}$.
Then, we obtain $\langle ReT \rangle \simeq 2.2$ and
$m_{3/2} \simeq 25$ TeV.
A height of bump at $ReT \sim 2.4$ 
in this example is $3m_{3/2}^2$,
because we uplifted the potential by $3m_{3/2}^2$.
In general, the height of bump is estimated as 
${\mathcal O}(m^2_{3/2})$.

\section{Moduli mixing in gauge coupling}

In several string models, the gauge kinetic function 
is obtained as a linear combination of 
two or more moduli fields.
In this section we show concrete examples.

First, 
in heterotic (M-)theory 
one-loop gauge coupling is given by \cite{Buchbinder:2003pi}
\beq
\nn
f_{strong}&=&S-\beta T +f_{M5},\\
\beta &\sim& 
\frac{1}{16\pi^4}\int_{CY}J\wedge \left[{\rm
    Tr}(F^{(2)})^2-\frac{1}{2}{\rm Tr}(R^2) \right],
\eeq
where $f_{strong}$ is the gauge kinetic function of a strong gauge
group and
$J$ is the K\"ahler form on CY with $h_{1,1}=1$.  
This can be seen from  10D Green-Schwarz term
$\int_{M_{10}} B_2 \wedge X_8$ or
11D Chern-Simmons term $\int_{M_{11}} C_3 \wedge G_4 \wedge
G_4$.
The third term $f_{M5}$ denotes a contribution from M5-brane position moduli 
$Y$ in the orbifold interval, 
such as $f_{M5} = \alpha Y^2/T,~\alpha \sim \int_{CY}J \wedge *_6 J$.
Hence, gluino condensation may generate moduli mixing superpotential.
\beq
W_{GC} \sim \exp[-a f_{strong}], ~~~ a=8\pi^2/N~~~{\rm for} ~~SU(N).
\eeq
That is, the fact that moduli mix in the gauge couplings, 
implies that moduli may mix in non-perturbative superpotential.
\footnote{
Moreover, M2-brane stretched between M5-brane and strong coupling M9-brane 
can generates \cite{Moore:2000fs}
\beq
\nn
W_{M2} \sim \exp\left[ -(\beta T-Y)\right].
\eeq
This is also moduli mixing superpotential. }

Type II models such as intersecting D-brane models or
magnetized D-brane models
have gauge couplings similar to those 
in heterotic models \cite{Lust:2004cx}. 
For example, in supersymmetric type IIB magnetized D-brane models on
$T^6/(Z_2 \times Z_2)$ orientifold with $h_{1,1}^{bulk}=1$
\footnote{Actually, $T^6/(Z_2 \times Z_2)$ orbifold, whose 
orbifold twists are 
\beq
\nn
\theta ~~:~~(z^1,~z^2,~z^3) &\rightarrow & (-z^1,~-z^2,~z^3)\\
\nn
\omega~~:~~(z^1,~z^2~,z^3)&\rightarrow & (z^1,~-z^2,~-z^3),
\eeq
have three K\"ahler forms in the bulk, that is $h_{1,1}^{bulk}=3$.
We identify indices of those cycles for simplicity. 
}, 
gauge couplings are given as follows,
\beq
\nn
 f_{mD7}&=&|m_{7}|S+|w_{7}|T ,
\\
f_{mD9}&=&m_9 S-w_9 T\qquad {\rm for}~~~O3/O7~{\rm system},
\label{mgauge}
\eeq
where $m_p,~w_p~(p=7,9)~\in {\mathbf Z}$
are Abelian magnetic flux contribution $F$ from world volume and
Wess-Zumino term.
In this case, Abelian gauge magnetic flux $F$ is 
quantized on a compact 2-cycle $C_2$ as $\int_{C_2}F \in {\mathbf Z}$. 
Then, $m_p~(p=7,9)$ is given by
$m_{7}= \int_{mD7} F \wedge F$ and
$m_9 = \int_{mD9} F\ \wedge F \wedge F $. 
On the other hand, $w_{9}$ is the winding number
on a wrapping 4-cycle 
and magnetic flux contribution, $w_9=\int_{mD9}*_6 
{J}_{bulk} \wedge F$ up to a numerical factor.
Moreover,  $w_7$ is the winding number of D7-brane on the 4-cycle.
Thus, signs of $m_9$ and $w_9$ can
depend on magnetic fluxes and SUSY conditions.
For example in \cite{Marchesano:2004xz},  one can find negative $m_9$
and $w_9$.
In addition T-duality action can exchange the winding number for the magnetic
number, but the result is similar, that is
\beq
 f_{mD9}&=&W_9 S-M_9 T \qquad {\rm for}~~~O9/O5~{\rm system},
\eeq 
where $W_9,~M_9~\in {\mathbf Z}$,
$W_9$ is the winding number on the 6-cycle and
$M_9$ is the winding number on the 2-cycle and magnetic flux contributions, 
$M_{9}=\int_{mD9} {J}_{bulk} \wedge F \wedge F$. 
Here we have neglected numerical factors again.

The gauge coupling on these magnetized brane is
written by
\beq
\frac{1}{g^2_{mD9}}= |m_9 ReS- w_9 ReT | .
\eeq 
Note that magnetic fluxes can contribute to RR tadpole condition of 
4-form potential and 8-form potential
\cite{Marchesano:2004xz}
\cite{Cascales:2003zp} and they should satisfy 
the tadpole cancellation condition,
\beq
\nn
\frac{N_{{\rm flux}}}{2}+
\sum _{p=7,9} \, \sum_{a={\rm stacks}}N^a_{p} m^a_p +Q^{others}_{3} =0,\\
\sum_{p=7,9} \, \sum_{a={\rm stacks}} N_p^a w_p^a +Q^{others}_{7}=0,
\eeq
where $Q^{others}_{p}~(p=3,7)$ are contributions of non-magnetized Dp-brane
and Op-plane and    
$N_p^a$ is the stack number of magnetized Dp-branes.
In this paper, we  treat $w_p,~m_p$ as free parameters
because we only concentrate on a stack of magnetized D-branes.

In Type IIA intersecting D-brane models, which are T-duals of the
above IIB string models, 
the above expressions of K\"ahler moduli change for complex
structure moduli. 
However, since there are 3-form and even-form fluxes, 
all geometric moduli can be frozen in this type IIA model
at low energy as a supersymmetric AdS vacuum.

In orbifold string theory, moduli in twisted sector, 
the so-called twisted moduli $M$, can exist \cite{Aldazabal:1998mr}. 
These modes
can contribute to gauge kinetic function on D-brane near orbifold fixed point, 
\beq
f_{Dp}= (S ~{\rm or}~T) + \sigma M ,
\eeq 
where $\sigma$ is ${\mathcal O}(0.1)-O(1)$ of parameter, depending on gauge and
orbifold group.
These twisted moduli may stabilize easily due to their K\"ahler
potential \cite{Abel:2000tf}, but
make little contribution to gauge coupling because the moduli are
related to orbifold collapsed cycle. 
Then, we may naturally have $\langle M \rangle \ll 1$.

\section{KKLT models with moduli-mixing superpotential}

As seen in the previous section, gauge kinetic 
functions are linear combinations of two or more moduli 
fields in several string models.
Thus, in this section we study the model that the 
gauge kinetic function for gluino condensation 
superpotential is a linear combination of two moduli, 
say $S$ and $T$, but one of them, $S$, is frozen 
around $M_{string}$, e.g. by flux compactification.\footnote{
In Ref.~\cite{AHK}, we will study the model that both of $S$ and $T$ remain 
light.}
For concreteness, we use the terminology of models 
based on O3/O7 type IIB string theory, in particular 
the gauge kinetic functions (\ref{mgauge}).
However, if we can realize the above situation in 
other string models, the following discussions are 
applicable for such string models.

After dilaton and complex structure moduli are frozen out 
in the ISD flux background around $M_{string}$,
we consider the following K\"ahler potential and superpotential 
at low energy,
\beq
K&=& -n_S \ln(\langle S \rangle +\langle \overline{S} \rangle  )
-n_T \ln(T+\overline{T})
,\quad n_S,~n_T \in \mathbf{N}, \\
W(T)&=&w(T) \pm Be^{\pm bT}, \qquad b>0 .
\label{models-W}
\eeq
For the first term $w(T)$ in R.H.S. of (\ref{models-W}), 
we study the following three cases 
\beq
w(T)&=& \langle \int G_3 \wedge \Omega \rangle, 
\eeq
\beq
w(T) &=&
e^{-\frac{8\pi^2 \langle S \rangle}{N_a}}, 
\eeq
or
\beq
w(T) &=&Ae^{-\frac{8\pi^2 T}{N_a}}.
\eeq
In the second case, the exponential term of $\langle S \rangle$ 
can be generated by gluino condensation
on a stack of $N_a$ D3-branes which is far from warped throat
\footnote{If we have another inflationary warped throat, such as $a_0' \sim 10^{-
3}$,
gluino condensation on a stack of $N$ D7-branes or D3-brane on the tip of that 
throat can also 
generate a $\langle S \rangle$ dependent superpotential
$w_0 \sim (a_0')^3 e^{-\frac{8\pi^2 \langle S \rangle}{N}}$.}.
At any rate, the first and second cases correspond to 
$w(T)=w_0=$ constant.
Thus, their potential analysis is the almost same.
The third case  
can be realized on a stack of $N_a$ non-magnetized D7-branes
or Euclidean D3-brane which is far from the throat.
Alternatively, this term can be generated by magnetized 
D7-branes, and in this case we have $A=e^{-8 \pi^2\langle S \rangle /N_a}$.
Next, we explain the second term $Be^{\pm bT}$ in
(\ref{models-W}).
Since  gauge coupling on magnetized D-brane is given by
$f_b \equiv m_bS \pm  w_bT$~
$(m_b,~w_b~\in {\mathbf N})$ like (\ref{mgauge}), 
strong dynamics on those branes may generate
a term like $e^{-cf}$.
Thus, the term $e^{-bT}$ ($b>0$) means that the superpotential 
can be generated by
gluino condensation on (non-)magnetized D7-brane by ({\ref{mgauge}}).
On the other hand, 
the term $e^{bT}$ can imply gluino condensation on magnetized D9-brane
and we may use this potential so far as a gauge coupling is weak,
i.e. 
\beq
m_b ReS  > w_b ReT.
\label{critical}
\eeq
In this case from (\ref{scales}) and (\ref{critical}), we need the condition
$\frac{m_b}{w_b}> \frac{\langle ReT \rangle}{\langle ReS \rangle}>1$.
Now, since the dilaton $S$ is stabilized at very high energy
because of 3-form fluxes, 
then we can write $f=m_bS \pm  w_bT \rightarrow m_b \langle S \rangle
\pm w_bT$.
Thus, 
at low energy, 
gluino condensation on these 
magnetized D-branes can generate a very suppressed value of $|B|$, 
which is given by 
\beq
B \equiv Ce^{-m_bc\langle S \rangle }&,& \quad c=\frac{8\pi^2}{N_b}~~{\rm
  for}~SU(N_b),
\eeq
i.e., 
$|B| \ll 1$.
We assume that $C$ is of ${\mathcal O}(1)$ and one may find $b=cw_b$.

Then, at the final stage we add the following uplifting potential,
\beq
V_L &=&\frac{D}{(T+\overline{T})^{n_p}},
\label{uplift-V}
\eeq
and the total potential is written as 
$V = V_F+V_L$.
We tune it such that $V=V_F+V_L=0$.
This uplifting potential for $n_p=2$ may be induced by adding a few
$\overline{{\rm D3}}$-branes at the tip of the throat.
If there is a magnetized D9-brane,  
presence of a few $\overline{{\rm D3}}$-branes
may give non-trivial effects to the D9-brane.
In such case, we  assume that the compact space is orbifolded,  
D3-branes are located at the orbifold fixed points in the bulk 
and {$\overline{{\rm D3}}$}-branes are located at the fixed point on the tip
of a warped throat
like \cite{Aldazabal:2000sa}. Then
the position of D3-branes are fixed and 
mass of open string tachyon may be nearly a TeV scale. 
Hence it may not affect to our model at this scale.
In this case, we can have exotic matter on $\overline{{\rm D3}}$-brane at a
TeV scale. 
Furthermore, we  neglect 
twisted moduli contribution to gauge coupling as 
$\langle M \rangle \ll 1$. 
We also assume that we
have a larger amount of total number of D3-branes and D7-branes in the bulk
than D9-branes,
in order not to change the geometry of \cite{Giddings:2001yu}.

Following \cite{Choi:2005ge},
we consider arbitrary integer of $n_p$ to study generic case.
At any rate, our models are well-defined as 4D effective 
supergravity models.
In what follows, we study four types of models 
corresponding to all of possibilities mentioned above.

\subsection{Model 1}

We study the following superpotential,
\beq
W&=& w_0 -B e^{-bT}.
\eeq
In this model, the second term is generated by 
gluino condensation on a magnetized D7-brane, where 
a gauge kinetic function is $ f=m_b\langle S \rangle +w_b T$.
If $|B| \gg |w_0|$, then this is similar to the previous KKLT
model. 
Thus, the modulus $T$ is stabilized at,
\beq
bReT \simeq \ln \left[
\frac{|B|}{|w_0|}
\right] \simeq -\frac{8\pi^2 m_b}{N_b}\langle S\rangle - \ln|w_0|.
\eeq
Therefore, we obtain
\beq
\frac{8\pi^2}{N_b}\langle Ref \rangle
\simeq
 \ln \left[ \frac{M_p}{m_{3/2}} \right],
\eeq
where we have used $B= Ce^{-8\pi^2 m_b\langle S \rangle/N_b}$, $C=N_b$,
$b=8\pi^2 w_b/N_b$, and 
the gravitino mass is obtained as $m_{3/2} \simeq w_0$.

When we add the uplifting potential (\ref{uplift-V}) 
and tune it such that $V=V_F+V_L=0$, 
then SUSY is broken and $F^T$ is induced as 
\beq
 \frac{F^T}{(T+\overline{T})} & \simeq &
 n_p \frac{3}{n_T}\frac{m_{3/2}}{b(T+\overline{T})} e^{-i\theta_W} .
\eeq
Thus, compared with the results in subsection \ref{KKLT},
$F^T$ becomes larger by the factor 
$\ln \frac{M_p}{|B|}=\frac{8\pi^2 m_b\langle S \rangle}{N_b}$
, and the modulus mass becomes 
lighter by the factor $\ln \frac{|B|}{M_p}$, i.e., 
$m_T = b(T + \overline T)m_{3/2}$.
Thus, we obtain 
\beq
\alpha \simeq \frac{{2n_T}}
{ 3n_p\left({\frac{m_b\langle ReS \rangle}{w_b \langle ReT
    \rangle}+1} \right)}.
\eeq
With $n_T=3,~n_p=2$, we have $0 < \alpha <1$. 
Hence, the modulus mediation and anomaly mediation 
are still comparable except the case with $|\alpha| \ll 1$, 
where the modulus mediation is dominant.

When $|B|(n_T+b(~\langle T+\overline{T} \rangle~)) \leq n_T |w_0|$, 
this analysis may not be  reliable in perturbative description, 
because of $\langle ReT \rangle \leq 0$. 
We may need quantum or
$\alpha'$ correction to K\"ahler potential
\cite{Becker:2002nn}. 
However, for simplicity, 
we do not consider such case here.

When the constant term $w_0$ is generated by fluxes, 
we have to fine-tune fluxes as 
$G_{(0,3)}/G_{(2,1)} \sim 10^{-13}$, 
in order to realize soft masses at the weak scale.
On the other hand, when the constant $w_0$ is generated as 
$w_0 = e^{-8 \pi^2\langle S \rangle /N_a}$, 
we do not need such fine-tuning for $\langle S \rangle ={\mathcal O}(1)$.
(See also Ref.~\cite{Loaiza-Brito:2005fa}.)

Generic form of potential in this model is similar to 
Figure \ref{fig1}.
The height of the bump is of ${\mathcal O}(m^2_{3/2})$, and 
this potential has the runaway behavior at the right 
of the bump.

\subsection{Model 2}

We study the following superpotential,
\beq
W&=& Ae^{-aT} -B e^{-bT}.
\eeq
This superpotential can be generated by 
 gluino condensation on magnetized D7-brane and
non-magnetized D7-brane.
We assume that $G_{(0,3)}=0$, then $w_0=0$ for simplicity. 
However,  the following results can change 
quantitatively but not qualitatively for the case with $w_0 \neq 0$
\cite{Choi:2005ge}. 
This model is called the racetrack model \cite{Krasnikov:1987jj}. 
For $a ReT,~bReT ={\mathcal O}(4\pi^2)$, 
the SUSY vacuum leads to 
\beq
\frac{Ae^{-aT}}{Be^{-bT}} &\simeq & \frac{b}{a} = 
\frac{w_bN_a}{N_b} \in {\mathbf R},
\\
\nn
ReT &\simeq &
\frac{1}{a-b}\ln \left [
\frac{|A|a}{|B|b}
\right], \\
&\simeq &
\frac{m_bN_a}{N_b-w_b N_a} \langle ReS \rangle ,\\
\nn
ImT &=& \frac{Arg(A)-Arg(B)}{a-b}+\frac{2\pi}{a-b}n,~~~n \in {\mathbf R},
\\
&=&-\frac{m_bN_a}{N_b-w_bN_a}\langle ImS \rangle + 
\frac{N_a N_b}{2(N_b-w_bN_a)}n ,
\\
m_T& \simeq & \frac{ab(T+\overline{T})^2}{n_T}m_{3/2}.
\eeq
Here, for simplicity,
we have assumed that $ReS \sim 1 $, 
$a=8\pi^2/N_a,~b= 8\pi^2 w_b/N_b,~A={N_a},~B=N_be^{-b m_b \langle S
  \rangle/w_b}$ and $w_b={\mathcal O}(1)$.
{}From (\ref{scales})
we can find the physical scales 
\beq
\nn
\frac{M_p}{M_{string}}= 2{\sqrt{\pi}}{\langle ReS \rangle}
\left(\frac{N_a m_b}{N_b-w_bN_a}\right)^{3/4}, \quad
\frac{M_{string}}{M_{KK}}= \left(\frac{N_a m_b}{N_b-w_bN_a}\right)^{1/4}.
\eeq
We require the  condition that the compactification
scale is smaller than the string scale, i.e.,
\beq
w_bN_a< {N_b} < (m_b+w_b)N_a.
\label{N_b}
\eeq
Then, we obtain 
\beq
& & \frac{W_{TT}}{W_T} \simeq -ab\frac{(T+\overline{T})}{n_T}, \\
& & \langle Ref \rangle = m_b\langle ReS \rangle + w_b \langle ReT \rangle \simeq 
\frac{m_bN_b}{N_b-w_bN_a}\langle ReS \rangle. 
\eeq
The gravitino mass is estimated as 
\beq
m_{3/2} \simeq e^{-a \langle ReT \rangle}.
\eeq

When we add the uplifting potential (\ref{uplift-V}) 
and tune it such that $V=V_F+V_L=0$, 
then SUSY is broken.
Generic form of potential is similar to Figure \ref{fig1}.
By generic analysis in Appendix {\ref{comp}}, 
we obtain 
\beq
\left|\frac{F^{T}}{(T+\overline{T})}\right|
&\simeq&3n_p \frac{m_{3/2}}{ab(T+\overline{T})^2}.
\eeq
Since we have 
\beq
\ln\left[
\frac{M_p}{m_{3/2}}
\right]
&\simeq &
a \langle ReT \rangle
= \frac{8\pi^2}{N_b}\langle Ref \rangle
=
\frac{8\pi^2 m_b}{N_b-w_bN_a} \langle ReS \rangle ,
\eeq
we estimate the ratio parameter,
\beq
\alpha &\simeq &
\frac{2}{n_p} \frac{b(T+\overline{T})}{3}
\sim  {\mathcal O}(4\pi^2).
\eeq
That implies that anomaly mediation is dominant in SUSY breaking.
These results have already been pointed out in \cite{Choi:2005ge}, 
but in our model we do not need to tune fluxes 
in order to obtain $ReT \sim 1$
so far as satisfying (\ref{N_b}) 
and $ReS \sim 1$.
Furthermore, since in this case we have
\beq
|W| \sim |C| \exp\left[-\frac{8\pi^2 m_b}{N_b-w_bN_a}\langle ReS
  \rangle \right] ,
\eeq
a very small scale may be generated by a value of $ReS$.
That is different from the usual racetrack model.

When we consider the case that both terms in the superpotential 
are generated by magnetized D7-branes, 
and replace as $A \rightarrow {N_a}
e^{-8\pi^2 m_a \langle S
  \rangle /N_a}$ with $a=8\pi^2w_a /N_a$, then we obtain  
\beq
\langle ReT \rangle 
\simeq \frac{(N_a m_b-N_bm_a)}{(w_a N_b - w_b N_a)}\langle ReS
\rangle ,
\eeq 
and
\beq
\nn
 \langle Ref_a \rangle &=& m_a \langle ReS\rangle +w_a \langle ReT
\rangle=N_{a}\cdot\frac{(w_am_b-m_aw_b)}{(w_a N_b - w_b N_a)}\langle
ReS\rangle ,\\
&=&
 \langle Ref_b \rangle \cdot \frac{N_a}{N_b} .
\eeq
In this case, the gravitino mass is estimated as 
\beq
m_{3/2} \simeq e^{-8\pi^2 \langle Ref_a \rangle/N_a},
\eeq
and the parameter $\alpha $ is estimated as  
\beq
\alpha \simeq \frac{\frac{16\pi^2}{N_b}w_b(T+\overline{T})}{3n_p\left(\frac{
m_a \langle
    ReS \rangle}
{w_a \langle ReT \rangle}+1 \right)}  .
\eeq
Its natural order is of ${\mathcal O }(4\pi^2)$.
That is, the anomaly mediation is dominant.
In a special parameter region, we may have $\alpha = {\mathcal O }(1)$, 
where the modulus mediation and anomaly mediation are comparable.

\subsection{Model 3}

We study the following superpotential,
\beq
W=w_0 +Be^{bT}.
\eeq
Now we consider the case that the gauge kinetic function 
is written by $f_b=m_b \langle S \rangle -w_b T$, and 
the condition of (\ref{critical}) should be satisfied.
In this case, we assume the presence of gluino condensation on a
magnetized D9-brane. 
For $b ReT \gg 1$, the modulus $T$ is stabilized at
\beq
b ReT &=& \ln \left[
\frac{|w_0| n_T}{|B|\left(b(T+\overline{T})-n_T \right)}
\right]
\simeq
\ln \left[
\frac{|w_0|}{|B|}
\right],\\
b ImT &=& Arg(w_0)-Arg(B)  ,
\eeq
where we have used $b=8\pi^2 w_b/N_b,~B=Ce^{-bm_b\langle S \rangle /w_b},~
C=N_b$. 
One might think that for $|w_0| \gg |B|$, 
we do not need to tune fluxes 
to realize $ReT \sim 1$ when $m_b \sim w_b$ and $ReS \sim 1$. 
However, 
in order to obtain a weak coupling on the
magnetized D-brane at the  cut-off scale, 
we need the following condition, 
\beq
\langle Ref_b \rangle=  {m_b} \langle ReS \rangle -w_b ReT
\simeq 
 -\frac{N_b}{8\pi^2}
\ln |w_0|  
\sim 1.
\label{model3-f}
\eeq
Then, we must tune the parameter as $|w_0| \sim \exp[-8\pi^2/N_b]$.
In the case with $w_0= \langle \, \int G_3 \wedge \Omega \, \rangle $,
we have to fine-tune fluxes.
However for $w_0=
e^{-8\pi^2 \langle S \rangle/N_a}$ with $G_{(0,3)}=0$, 
we do not need such fine-tuning.
In this model, since moduli can be stabilized as $w_0 \sim \langle
Be^{bT} \rangle$,
the gravitino mass is given by $m_{3/2} \sim |w_0|$.
For $m_b>1$, we can find 
\beq
\frac{8\pi^2}{N_b}\langle Ref_b \rangle \simeq \ln \left[ 
\frac{M_p}{m_{3/2}}
\right] ,
\eeq
from (\ref{model3-f}).

When we add the uplifting potential (\ref{uplift-V}) 
and tune it such that $V=V_F+V_L=0$, 
SUSY is broken and $F^T$ is induced.
Since $W_{TT}= +bW_T$
we obtain 
\beq
& & \frac{F^T}{(T+\overline{T})}  \simeq 
 -n_p \frac{3}{n_T}\frac{m_{3/2}}{b(T+\overline{T})} e^{-i\theta_W} ,\\
& & \alpha \simeq
-\frac{2n_T w_b ReT}{3n_p \langle Ref_b \rangle}
=-\frac{2n_T }{3n_p \left(
\frac{m_b \langle ReS \rangle}{w_b \langle ReT \rangle}-1\right)}.
\eeq
That is, the ratio parameter $\alpha $ is negative.
Naturally we would
have $|\alpha|={\mathcal O}(1)$.

This potential does not seem to have the runway behavior.
That has important implications on cosmology as will 
discussed in the next section.
These properties are remarkably different from the KKLT model, 
although this model only changes to $w_b \rightarrow -w_b$ from model 1.

\subsection{Model 4}

We study the following superpotential,
\beq
W=Ae^{-aT}+Be^{bT} .
\eeq
In this model, we assume that non-perturbative effects 
 on the  magnetized D9-brane 
and the non-magnetized D7-brane generate the above superpotential.
We also assume vanishing 3-form flux $G_{(0,3)}=0$.
In this model, the  condition (\ref{critical}) should be satisfied.
This superpotential can be obtained also in heterotic M-theory 
\cite{Becker:2004gw}, where 
the first term can be originated from membrane instanton 
and the second term is originated from gluino 
condensation on the strong coupled fixed plane\footnote{
See also \cite{Curio:2001qi}.
}.
In this heterotic model
the orbifold interval $T$ can be
stabilized, because of signs of exponents.

This type IIB model is the same as the model 2 except  using
magnetized D9-brane.
For $a ReT,~bReT \gg 1$, the modulus $T$ is stabilized as 
\beq
ReT &\simeq&
\frac{1}{a+b}\ln \left[
\frac{|A|a}{|B|b}
\right]
\simeq
\frac{N_a m_b}{N_b+w_bN_a}\langle ReS \rangle , \\
ImT&=&\frac{Arg(A)-Arg(B)}{a+b}
= - \frac{N_a m_b}{N_b+w_bN_a}\langle ImS \rangle ,
\eeq
where we have assumed the same parameters as model 2.
One may find that $ReT$ tends to be smaller than model 2, 
and  we may 
need large magnetic fluxes $m_b$.
{}From (\ref{scales})
we can find the physical scales
\beq
\nn
\frac{M_p}{M_{string}}= 2{\sqrt{\pi}}{\langle ReS \rangle}
\left(\frac{N_a m_b}{N_b+w_bN_a}\right)^{3/4}, \quad
\frac{M_{string}}{M_{KK}}= \left(\frac{N_a m_b}{N_b+w_bN_a}\right)^{1/4} .
\eeq
Thus, we require the following condition like model 2,
\beq
0<N_b < N_a (m_b-w_b).
\label{N_b-2}
\eeq
We have the gauge kinetic function on the magnetized D9-brane
\beq
\langle Ref \rangle 
= \frac{m_bN_b}{N_b +w_bN_a} \langle Re S \rangle .
\eeq
Hence, the gauge coupling on the magnetized D9-brane can be weak 
 so far as satisfying (\ref{N_b-2}) and $\langle ReS \rangle \sim 1
 $.
Since the dynamical scale on this D-brane 
is obtained as  $ |C| \exp[-8\pi^2 m_b\langle ReS \rangle
/(N_b+ w_bN_a)] \ll 1$, we can generate a small scale as model 2.
Furthermore, 
without fine-tuning, we can have moderate values of $\langle ReT \rangle$ and
the gauge coupling
so far as satisfying (\ref{N_b-2}).
Other properties are obtained from model 2 by replacing 
$w_b \rightarrow -w_b$.
For example we have negative $\alpha$.

When we replace $A \rightarrow {N_a}
e^{-8\pi^2 m_a \langle S \rangle /N_a}$ with $a=8\pi^2 w_a/N_a$, 
results are also obtained from model 2 by replacing 
$w_b \rightarrow -w_b$.
For example, we obtain the parameter $\alpha$,
\beq
\alpha \simeq - \frac{\frac{16\pi^2}{N_b}w_b(T+\overline{T})}{3n_p\left(\frac{
m_a \langle
    ReS \rangle}
{w_a \langle ReT \rangle}-1 \right)} .
\eeq
It is negative and its natural order is ${\mathcal O}(4 \pi^2)$.

\begin{figure}
\epsfxsize=0.7\textwidth
\centerline{\epsfbox{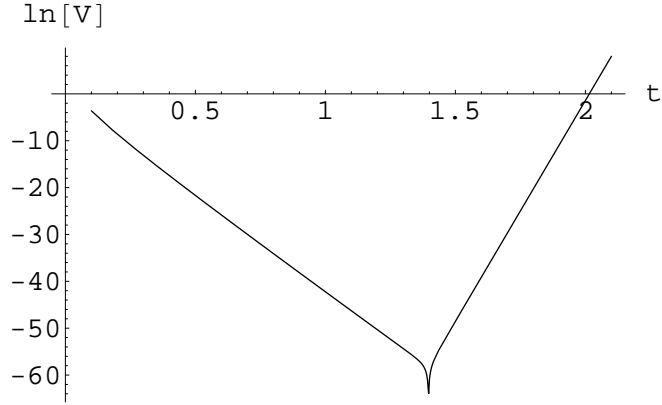}}
\caption{The potential of model 4
with parameters, $n_T=3,~n_S=1,~ n_p=2$,  
$N_a=4,~ N_b=5,~ m_b=6,~ w_b= 3,~\langle ReS \rangle =1,~ D= 2.2
\times 10^{-27}$.
Then, we obtain $\langle ReT \rangle \simeq 1.4$, 
$M_{string}/M_p \simeq 0.22$ and $m_{3/2} \simeq 40$ TeV.
A horizontal axis is $t=ReT$ and a vertical axis is
$\ln[V]$. This potential can make sense until $ReT < m_b 
\langle ReS \rangle /w_b = 2$.}
\label{fig2}
\end{figure}

Here we show an example of potential of model 4 in Figure \ref{fig2}.
It is quite different from model 1 and model 2, while 
in model 3 a similar potential is obtained.
That has cosmologically important implications, 
as will be discussed in the next section.


\section{Implications on SUSY phenomenology and 
cosmology}

In this section, we discuss implications of our results on 
SUSY phenomenology and cosmology.

Concerned about SUSY breaking, our results are 
qualitatively similar to those in Ref.~\cite{Choi:2005ge}, 
that is, in model 1 and 3, the modulus mediation and 
anomaly mediation are comparable, while 
in model 2 and 4, the anomaly mediation is dominant.
Our models generalize those results.
Since the parameters $A$ and $B$ can expand wider range
in our models, 
the parameter $\alpha$ varies a wider parameter region, 
even if $n_T$ and $n_p$ are fixed.
In model 3, we obtain a negative value of $\alpha$, which would be
naturally of ${\mathcal O}(1)$. In model 1, we have $0< \alpha <1$
when we fix $n_T=3$ and $n_p =2$. In addition, a value of $\alpha$ in
model 2 and 4 is of ${\mathcal O}(4\pi^2)$. 
However, whether the modulus mediation and anomaly mediation 
are additive or destructive depends on gauge kinetic functions 
of visible sector.
Suppose that the gauge kinetic function of the visible sector 
is given as
\beq
f_v = m_v S + w_v T,
\eeq
where $w_v$ can be positive and negative.
Then, whether the modulus mediation and the anomaly 
mediation are additive or destructive depends on 
the signs of $\alpha $ and $w_v$.
Furthermore, the size of gaugino masses 
induced by $F^T$ is written as
\beq
M_{1/2}^{(T)}=\frac{F^T}{T+\overline{T}}
\frac{1}{\left(\frac{m_v \langle ReS \rangle}{w_v \langle ReT
      \rangle}+1\right)}.
\eeq
Thus, it can be enhanced or suppressed.
At any rate, soft SUSY breaking terms in our models have 
a rich structure.
We will study their spectra and phenomenological aspects 
elsewhere.

Next, we discuss cosmological aspects of our models.
The anomaly mediation is sizable except the special case with 
$|\alpha | \ll 1$.
That implies that soft masses in the visible sector are 
suppressed by one-loop factor compared with the 
gravitino mass $m_{3/2}$, that is, 
the gravitino can be heavier like $m_{3/2} = {\mathcal O}(10)$ TeV.
The modulus can have much larger mass.
Therefore, we may avoid the gravitino problem and 
the moduli problem in all of models like 
the usual KKLT 
model \cite{Choi:2005ge,Choi:2005uz,Endo:2005uy,Falkowski:2005ck}.
Furthermore, the potential forms of model 3 and 4 have more interesting 
aspects as discussed below.

The potential of model 1 and 2 has the runaway behavior at 
the right region of the bump like Figure \ref{fig1}, 
and the height of the bump is of ${\mathcal O}(m^2_{3/2})$.
When an initial value of $T$ is in the right of the bump, 
$T$ goes to infinity.
Furthermore, when an initial value of $T$ is quite small
$T \ll 1$, $T$ overshoots the favorable 
minimum and goes to infinity.
Thus, we have to fine-tune the initial condition 
such that $T$ is trapped at the favorable minimum.
That is the overshooting problem \cite{Brustein:1992nk}.

This type of potential has another problem, 
that is, destabilization due to finite temperature effect.
The finite temperature effect induces the additional 
potential term \cite{kapusta}, 
\beq
\Delta V = (\alpha_0 + \alpha_2 g^2) \widehat T^4,
\eeq
where $\widehat T$ denotes the temperature.
The coefficients, $\alpha_0$ and $\alpha_2$,  
are written by group factors of massless modes, 
and in most of case $\alpha_2$ is positive.
For example, we have $\alpha_2=\frac{3}{8\pi^2}(N_c^2-1)(N_c+3N_f)$ 
for SUSY $SU(N_c)$ gauge theory with $N_f$ flavors of matter 
multiplets.
In models 1 and 2, the gauge kinetic function is 
obtained as $f=mS+wT$.
Thus, the potential term due to finite temperature 
effect is written as 
\beq
\Delta V = \left[ \alpha_0 + \alpha_2 
\frac{1}{m\langle S \rangle + w T} \right] \widehat T^4 .
\eeq
This term destabilizes $T$ at not so high temperature, but 
the temperature corresponding to the intermediate scale 
\cite{Buchmuller:2004xr}.

The above two problems are not problems only for our models 1 and 2, 
but are rather generic problems.
On the other hand, the potential in models 3 and 4 has the term 
like $Be^{bT}$.
Such term may avoid the overshooting problem, and 
this term is reliable except when $Be^{bT} \geq {\mathcal O}(1)$, because 
in that region we would have uncontrollable effects.
However, this reliable region of the potential is 
much higher than ${\mathcal O}(m^2_{3/2})$, which is the height of 
the bump in model 1 and 2 as well as other potentials 
with this type of potential forms.
The same behavior of the potential is helpful to avoid 
the destabilization problem due to finite temperature effects.
For example, in model 3 the gauge kinetic function is written as 
$f = mS - wT$.
Thus, the potential term due to finite temperature effects 
is written as 
\beq
\Delta V = \left[ \alpha_0 + \alpha_2 
\frac{1}{m\langle S \rangle - w T} \right ]\widehat T^4 .
\eeq
That makes $T$ shift to a smaller value, 
because smaller $T$ corresponds to weaker coupling.
Therefore, the VEV of $T$ does not destabilize.
Model 4 also has the same behavior.
As results, the potential form in models 3 and 4 are cosmologically 
interesting from the viewpoint to avoid the 
overshooting problem and destabilization problem due to finite 
temperature.

\section{Conclusion}

We have considered the KKLT model with moduli-mixing superpotential,
assuming that
one of them is frozen.
Such superpotential can be obtained e.g. by gluino condensation 
on magnetized D-branes, while it may be generated in other setups. 
We have studied four types of models.
In these models, the hierarchy
between the Planck scale and gravitino mass can be written by gauge coupling,
 such as
$\ln(M_p/m_{3/2}) \simeq 8\pi^2 \langle Ref \rangle /N$, 
that is, magnetic
fluxes can generate a large hierarchy.
Model 1 is almost the same as the KKLT model, but the ratio 
$\alpha$ between anomaly
mediation and modulus mediation can take various values.
Models 2, 3, 4 do not require fine-tuning of 3-form fluxes 
to realize $\langle ReT
\rangle \sim 1$ because of very small coefficient 
$B$. 
However, model 3 needs fine-tuning of 3-form fluxes 
in order to obtain a weak coupling on 
the magnetized D9-brane, but in the case that $w_0$ can be generated by gluino
condensation on D3-brane, we may not need to tune minutely. 
We may need to tune slightly the open string sector such as magnetic
fluxes, the winding number and the number of D-branes instead of 3-form
fluxes (in the closed string sector).  
In model 3 and 4, $\alpha$ becomes negative.
All of models lead to a rich structure of SUSY breaking 
including new patterns of soft SUSY breaking terms.
Such spectra and their phenomenological aspects would 
be studied elsewhere.

In most of models except for $|\alpha| \ll 1$, the gravitino and 
moduli masses are of ${\mathcal O}(10)$ TeV or much heavier.
Such spectrum is important to avoid the gravitino problem 
and the moduli problem.
Furthermore, the potential form of models 3 and 4 
have good properties for cosmology because of 
the exponential factor, $\exp[+bT]$ for $b>0$.
That may avoid the overshooting problem and 
the destabilization problem due to finite temperature effects.

We have studied the models that two moduli fields $S$ and $T$ 
have mixing in superpotential, assuming one of them $S$ is frozen 
already around the string scale $M_{string}$.
Alternatively, both of them may remain light.
In Ref.~\cite{AHK}, we will study such models, i.e., moduli 
stabilization and SUSY breaking.

\section*{Acknowledgment}

T.~H.\/ is supported in part by the Grand-in-Aid for Scientific
Research \#171643.
T.~K.\/ is supported in part by the Grand-in-Aid for Scientific
Research \#16028211 and \#17540251.
H.~A.\/, T.~H.\/ and T.~K.\/ are supported in part by 
the Grant-in-Aid for
the 21st Century COE ``The Center for Diversity and
Universality in Physics'' from the Ministry of Education, Culture,
Sports, Science and Technology of Japan.

\appendix

\section{Analysis on the KKLT type scalar potential}
\label{comp}

In this appendix, we summarize analysis on 
the KKLT type scalar potential.


\subsection{SUSY potential}

First we consider the following F-term scalar potential 
$V_F$ of supergravity model,
\beq
V_F&=&e^{G}\left[
G_IG_{\overline{J}}G^{I\overline{J}}-3
\right],\\
G&=&K+\ln|W|^2,
\eeq 
where $G_{I}=\partial_I G$, $G_{I\overline{J}}=\partial_I
\partial_{\overline{J}}G$.

The first derivative of $V_F$ is obtained as 
\beq
\nn
\partial_{I} V_F 
&=&e^{G} \Bigr[
G_I \cdot \left( G_K G_{\overline{L}} G^{K \overline{L}}-2 \right)\\
& & ~~~~~~~~~~~~~~~+G_{\overline{L}}\cdot 
\left( G_{KI}G^{K \overline{L}} + G_K \left(\partial_I G^{K \overline{L}} \right) 
\right)
\Bigr],
\label{pv1}
\\
\nn
&=&
\nn
F^K\left[
(\partial_I K_{K\overline{L}})\overline{F}^{\overline{L}}+K_{K\overline{L}}
\partial_I\overline{F}^{\overline{L}}
\right]\\
& &~~~~~~~~~~~~~~~~+ \overline{F}^{\overline{L}}
\left[
K_{K\overline{L}}\partial_I F^K +3e^{K/2}\overline{W}K_{I\overline{L}}
\right],
\label{pv2}
\eeq
where we have defined $F^I=-K^{I\overline{J}}e^{K/2}\overline{D_{J}W}$. 
Thus, the SUSY point, i.e. $G_I=D_IW=0$, satisfies the 
stationary condition, $\partial_I V_F =0$. 
At such SUSY point, mass matrices are given by
\beq
V_{IJ}&=&-e^{G}G_{IJ},\\
V_{\overline{I}\overline{J}}&=&-e^{G}G_{\overline{I}\overline{J}},\\
V_{I\overline{J}}&=&e^{G}\left[
-2G_{I\overline{J}}+G_{IK}G_{\overline{J}\overline{L}}G^{K\overline{L}}
\right].
\eeq

Here we concentrate to the model with only single field $X$.
In this case, mass matrix is given as 
\beq
\left(
\begin{array}{cc}
 V_{X\overline{X}}+Re(V_{XX}) & -Im(V_{XX})  \\
 -Im(V_{XX}) &  V_{X\overline{X}}-Re(V_{XX})  
\end{array}
\right),
\eeq
in the basis of $(Re(x),Im(x))$, where 
$x=X-X_0$ with $X_0$ satisfying 
$\partial_X V|_{X=X_0}=0$.
Their eigenvalues are obtained as 
\be
V_{X\overline{X}} \pm |V_{XX}|.
\ee
Therefore, the SUSY point corresponds to the 
minimum of the potential if $V_{X\overline{X}} > |V_{XX}|$, 
that is, 
\beq
|G_{XX}|>2G_{X\overline{X}}.
\eeq

\subsection{KKLT type potential with uplifting potential}


Now we consider the KKLT type potential.
We use the following form of K\"ahler potential and 
generic form of superpotential, 
\beq
K&=&-n_T \ln(T+\overline{T})
, \qquad n_T \in {\mathbf N},\\
W&=&W(T) .
\eeq
The original KKLT model has $n_T = 3$. 
The SUSY point, i.e., $D_TW=0$, is 
obtained as 
\beq
(T+\overline{T})&=&n_T\frac{W}{W_T}~~\in
\mathbf{R}.
\label{vat}
\eeq
This point corresponds to the minimum of $V_F$ 
if the VEV of $T$ satisfies 
\beq
\left|\frac{n_T(n_T-1)}{(T+\overline{T})^2}
-\frac{W_{TT}}{W}\right|& >& \frac{2n_T}{(T+\overline{T})^2}.
\eeq 
We consider the case that the above condition is satisfied, 
but such minimum of $V_F$ is 
the SUSY AdS vacuum with the vacuum energy 
$V_F = -3 e^K|W|^2=-3m^2_{3/2}$.

In order to realize the dS vacuum, 
we add the following form of uplifting potential, 
\be
V_{L} = \frac{D}{(T+\overline{T})^{n_p}}, \qquad 0< D \ll 1 .
\ee
In the original KKLT model, this is originated from 
$\overline D3$-bane and we have $n_p=2$.
Here following Ref.~\cite{Choi:2004sx,Choi:2005ge}, we consider 
generic integer for $n_p$.
Now the total potential is written as 
\beq
V=V_{F}+V_{L},
\eeq
and adding $V_{L}$ changes only vacuum value of $ReT$ from (\ref{vat}) 
but not $ImT$.
We demand that a change of $ReT$ is small and 
the cosmological constant  $\langle V \rangle$
vanishes, that is,  
\beq 
\langle V_L \rangle =-\langle V_F \rangle \simeq  3m_{3/2}^2.
\label{<v>}
\eeq
{}From (\ref{pv2}), the first derivative of $V$ is written as 
\beq
\nn
\partial_T V&=&
F^T\left[
\frac{-2n_T}{(T+\overline{T})^3}\overline{F}^{\overline{T}}
+\frac{n_T}{(T+\overline{T})^2}\partial_T
\overline{F}^{\overline{T}}
\right]\\
\nn
& &+
\overline{F}^{\overline{T}}\left[
\frac{n_T}{(T+\overline{T})^2}\partial_T
F^T+3e^{K/2}\overline{W}\frac{n_T}{(T+\overline{T})^2}
\right]\\
\nn
&&~~~~-\frac{n_p}{(T+\overline{T})}V_L .
\label{lifp}
\eeq
Now, we expand as $T=T_{SUSY}+\delta T$ such as $D_TW|_{
T=T_{SUSY}}=0$.
Then, the F-term and its derivatives are evaluated as 
\beq
\overline{F}^{\overline{T}} &=& e^{K/2}\frac{(T+\overline{T})}{n_T}(n_TW-(T+
\overline{T})W_T),\\
\partial_T \overline{F}^{\overline{T}}
&\simeq&
a(T+\overline{T})\cdot e^{K/2}W=
a(T+\overline{T})m_{3/2},\\
\partial_{\overline{T}}\overline{F}^{\overline{T}}
&\simeq&-e^{K/2}W=-m_{3/2} .
\eeq
Here we have used $W_T \simeq n_{T}W/(T+\overline{T}) $ and
defined a real parameter $a$,
\be
a \equiv -\frac{W_{TT}}{W_T}.
\ee
We have also assumed that 
$|a| \cdot ReT_{SUSY} \gg 1$. 
There is  an important point for CP
phase.
Since 
\beq
\delta T, ~~~\frac{W_T}{W},~~~\frac{W_{TT}}{W}~
\in \mathbf{R},
\eeq
the CP phase of $W$ and $\overline{F}^{\overline{T}}$ 
are the same.
That is, 
when we write $W=|W|e^{i\theta_W}$, then 
we have $\overline{F}^{\overline{T}} \propto e^{i\theta_W}$. 
Using the above results, 
we can write
\beq
\partial_T V 
&\simeq&
m_{3/2}\frac{n_T}{(T+\overline{T})^2}\cdot \left[
a(T+\overline{T})F^T 
\right]-\frac{n_p}{(T+\overline{T})}V_L .
\eeq
Therefore, for $aReT \gg 1$ 
the stationary condition $\partial_T V = 0$ and 
the condition for the vanishing cosmological constant 
lead to 
\beq
 \frac{F^T}{(T+\overline{T})} & \simeq &
 n_p \frac{3}{n_T}\frac{m_{3/2}}{a(T+\overline{T})} e^{-i\theta_W} ,
\\
\frac{\delta T}{T_{SUSY}}&\simeq& \frac{n_p}{2}
\frac{3}{n_T}\frac{1}{a^2(ReT_{SUSY})^2}~~~ \ll 1.
\eeq

Furthermore, the F-component of conformal compensator superfield $\phi$
is given by 
\beq
\frac{F_{\phi}}{\phi_0}&=&e^{K/2}\overline{W} + \frac{1}{3}K_I F^I ,\\
&\simeq & e^{-i\theta_W}m_{3/2}, 
\eeq
where $\phi_0~ (\in {\mathbf R})$ is a scalar component of $\phi$.
Hence, CP phases of $F_{\phi}$ and $F^T$ are 
aligned \cite{Choi:2005ge,Choi:1993yd}. 
It is useful to define the ratio $\alpha$ as follows, 
\beq
\alpha \equiv \frac{F_{\phi}(T+\overline{T})}{\phi_0 \ln(M_p/ m_{3/2})
  F^T }
=\frac{n_T a (T+\bar{T}) }{ 3n_p\ln(M_p/ m_{3/2}) }, 
\eeq
because of 
$\ln(M_p/ m_{3/2}) \sim {\mathcal O} (4 \pi^2 )$.


\begin{thebibliography}{99}

\bibitem{Giddings:2001yu}
S.~B.~Giddings, S.~Kachru and J.~Polchinski,
Phys.\ Rev.\ D {\bf 66}, 106006 (2002)
[arXiv:hep-th/0105097].

\bibitem{DeWolfe:2005uu}
O.~DeWolfe, A.~Giryavets, S.~Kachru and W.~Taylor,
JHEP {\bf 0507}, 066 (2005)
[arXiv:hep-th/0505160].
%
J.~P.~Derendinger, C.~Kounnas, P.~M.~Petropoulos and F.~Zwirner,
Nucl.\ Phys.\ B {\bf 715}, 211 (2005)
[arXiv:hep-th/0411276].
%
G.~Villadoro and F.~Zwirner,
JHEP {\bf 0506}, 047 (2005)
[arXiv:hep-th/0503169].



\bibitem{Becker:2005nb}
K.~Becker and L.~S.~Tseng,
arXiv:hep-th/0509131.

\bibitem{Kachru:2003aw}
S.~Kachru, R.~Kallosh, A.~Linde and S.~P.~Trivedi,
Phys.\ Rev.\ D {\bf 68}, 046005 (2003)
[arXiv:hep-th/0301240].

\bibitem{Choi:2004sx}
  K.~Choi, A.~Falkowski, H.~P.~Nilles, M.~Olechowski and S.~Pokorski,
  JHEP {\bf 0411}, 076 (2004)
  [arXiv:hep-th/0411066].

\bibitem{Choi:2005ge}
K.~Choi, A.~Falkowski, H.~P.~Nilles and M.~Olechowski,
Nucl.\ Phys.\ B {\bf 718}, 113 (2005)
[arXiv:hep-th/0503216].

\bibitem{Randall:1998uk}
L.~Randall and R.~Sundrum,
Nucl.\ Phys.\ B {\bf 557}, 79 (1999)
[arXiv:hep-th/9810155].

\bibitem{Giudice:1998xp}
G.~F.~Giudice, M.~A.~Luty, H.~Murayama and R.~Rattazzi,
JHEP {\bf 9812}, 027 (1998)
[arXiv:hep-ph/9810442].

\bibitem{Choi:2005uz}
K.~Choi, K.~S.~Jeong and K.~i.~Okumura,
JHEP {\bf 0509}, 039 (2005)
[arXiv:hep-ph/0504037].

\bibitem{Endo:2005uy}
M.~Endo, M.~Yamaguchi and K.~Yoshioka,
Phys.\ Rev.\ D {\bf 72}, 015004 (2005)
[arXiv:hep-ph/0504036].

\bibitem{Falkowski:2005ck}
A.~Falkowski, O.~Lebedev and Y.~Mambrini,
arXiv:hep-ph/0507110.

\bibitem{Choi:2005hd}
K.~Choi, K.~S.~Jeong, T.~Kobayashi and K.~i.~Okumura,
arXiv:hep-ph/0508029; 
%
R.~Kitano and Y.~Nomura,
arXiv:hep-ph/0509039.

\bibitem{Choi:1985bz}
K.~Choi and J.~E.~Kim,
Phys.\ Lett.\ B {\bf 165}, 71 (1985).

\bibitem{Ibanez:1986xy}
L.~E.~Ibanez and H.~P.~Nilles,
Phys.\ Lett.\ B {\bf 169}, 354 (1986); 
%
J.~P.~Derendinger, L.~E.~Ibanez and H.~P.~Nilles,
Nucl.\ Phys.\ B {\bf 267}, 365 (1986); 
%
L.~J.~Dixon, V.~Kaplunovsky and J.~Louis,
Nucl.\ Phys.\ B {\bf 355}, 649 (1991).

\bibitem{Banks:1996ss}
T.~Banks and M.~Dine,
Nucl.\ Phys.\ B {\bf 479}, 173 (1996)
[arXiv:hep-th/9605136].

\bibitem{Choi:1997an}
K.~Choi,
Phys.\ Rev.\ D {\bf 56}, 6588 (1997)
[arXiv:hep-th/9706171].

\bibitem{Nilles:1997vk}
H.~P.~Nilles and S.~Stieberger,
Nucl.\ Phys.\ B {\bf 499}, 3 (1997)
[arXiv:hep-th/9702110].

\bibitem{Buchbinder:2003pi}
E.~I.~Buchbinder and B.~A.~Ovrut,
Phys.\ Rev.\ D {\bf 69}, 086010 (2004)
[arXiv:hep-th/0310112]; 
%
  A.~Lukas, B.~A.~Ovrut and D.~Waldram,
  Nucl.\ Phys.\ B {\bf 532}, 43 (1998)
  [arXiv:hep-th/9710208]; 
%
  A.~Lukas, B.~A.~Ovrut and D.~Waldram,
  Phys.\ Rev.\ D {\bf 57}, 7529 (1998)
  [arXiv:hep-th/9711197].


\bibitem{Curio:2000dw}
G.~Curio and A.~Krause,
Nucl.\ Phys.\ B {\bf 602}, 172 (2001)
[arXiv:hep-th/0012152];
%
G.~Curio and A.~Krause,
Nucl.\ Phys.\ B {\bf 693}, 195 (2004)
[arXiv:hep-th/0308202].

\bibitem{Cremades:2002te}
D.~Cremades, L.~E.~Ibanez and F.~Marchesano,
JHEP {\bf 0207}, 009 (2002)
[arXiv:hep-th/0201205].

\bibitem{Lust:2004cx}
D.~Lust, P.~Mayr, R.~Richter and S.~Stieberger,
Nucl.\ Phys.\ B {\bf 696}, 205 (2004)
[arXiv:hep-th/0404134].

%



\bibitem{AHK}
H.~Abe, T.~Higaki and T.~Kobayashi,
arXiv:hep-th/0512232.


\bibitem{Burgess:1997pj}
C.~P.~Burgess, A.~de la Macorra, I.~Maksymyk and F.~Quevedo,
Phys.\ Lett.\ B {\bf 410}, 181 (1997)
[arXiv:hep-th/9707062].

\bibitem{Kachru:2003sx}
S.~Kachru, R.~Kallosh, A.~Linde, J.~Maldacena, L.~McAllister and S.~P.~Trivedi,
JCAP {\bf 0310}, 013 (2003)
[arXiv:hep-th/0308055].

\bibitem{Klebanov:2000hb}
I.~R.~Klebanov and M.~J.~Strassler,
JHEP {\bf 0008}, 052 (2000)
[arXiv:hep-th/0007191].

\bibitem{Randall:1999ee}
L.~Randall and R.~Sundrum,
Phys.\ Rev.\ Lett.\  {\bf 83}, 3370 (1999)
[arXiv:hep-ph/9905221].

\bibitem{Gukov:1999ya}
S.~Gukov, C.~Vafa and E.~Witten,
Nucl.\ Phys.\ B {\bf 584}, 69 (2000)
[Erratum-ibid.\ B {\bf 608}, 477 (2001)]
[arXiv:hep-th/9906070]; 
%
T.~R.~Taylor and C.~Vafa,
Phys.\ Lett.\ B {\bf 474}, 130 (2000)
[arXiv:hep-th/9912152];
%
G.~Curio, A.~Klemm, D.~Lust and S.~Theisen,
Nucl.\ Phys.\ B {\bf 609}, 3 (2001)
[arXiv:hep-th/0012213]; 
%
O.~DeWolfe and S.~B.~Giddings,
Phys.\ Rev.\ D {\bf 67}, 066008 (2003)
[arXiv:hep-th/0208123].

\bibitem{Kachru:2002he}
S.~Kachru, M.~B.~Schulz and S.~Trivedi,
JHEP {\bf 0310}, 007 (2003)
[arXiv:hep-th/0201028]; 
%
P.~K.~Tripathy and S.~P.~Trivedi,
JHEP {\bf 0303}, 028 (2003)
[arXiv:hep-th/0301139];
%
A.~Giryavets, S.~Kachru, P.~K.~Tripathy and S.~P.~Trivedi,
JHEP {\bf 0404}, 003 (2004)
[arXiv:hep-th/0312104].


\bibitem{Camara:2003ku}
P.~G.~Camara, L.~E.~Ibanez and A.~M.~Uranga,
Nucl.\ Phys.\ B {\bf 689}, 195 (2004)
[arXiv:hep-th/0311241]; 
%
P.~G.~Camara, L.~E.~Ibanez and A.~M.~Uranga,
Nucl.\ Phys.\ B {\bf 708}, 268 (2005)
[arXiv:hep-th/0408036]; 
%
O.~Aharony, Y.~E.~Antebi and M.~Berkooz,
arXiv:hep-th/0508080.

\bibitem{Kachru:2002gs}
S.~Kachru, J.~Pearson and H.~L.~Verlinde,
JHEP {\bf 0206}, 021 (2002)
[arXiv:hep-th/0112197].






\bibitem{Denef:2004ze}
F.~Denef and M.~R.~Douglas,
JHEP {\bf 0405}, 072 (2004)
[arXiv:hep-th/0404116].


\bibitem{Gorlich:2004qm}
L.~Gorlich, S.~Kachru, P.~K.~Tripathy and S.~P.~Trivedi,
JHEP {\bf 0412}, 074 (2004)
[arXiv:hep-th/0407130].

\bibitem{Choi:1993yd}
K.~Choi,
Phys.\ Rev.\ Lett.\  {\bf 72}, 1592 (1994)
[arXiv:hep-ph/9311352].

\bibitem{Moore:2000fs}
G.~W.~Moore, G.~Peradze and N.~Saulina,
Nucl.\ Phys.\ B {\bf 607}, 117 (2001)
[arXiv:hep-th/0012104].

\bibitem{Marchesano:2004xz}
F.~Marchesano and G.~Shiu,
JHEP {\bf 0411}, 041 (2004)
[arXiv:hep-th/0409132].

\bibitem{Cascales:2003zp}
J.~F.~G.~Cascales and A.~M.~Uranga,
JHEP {\bf 0305}, 011 (2003)
[arXiv:hep-th/0303024].

\bibitem{Aldazabal:1998mr}
G.~Aldazabal, A.~Font, L.~E.~Ibanez and G.~Violero,
Nucl.\ Phys.\ B {\bf 536}, 29 (1998)
[arXiv:hep-th/9804026]; 
%
L.~E.~Ibanez, R.~Rabadan and A.~M.~Uranga,
Nucl.\ Phys.\ B {\bf 542}, 112 (1999)
[arXiv:hep-th/9808139].

\bibitem{Abel:2000tf}
S.~A.~Abel and G.~Servant,
Nucl.\ Phys.\ B {\bf 597}, 3 (2001)
arXiv:hep-th/0009089; 
%
T.~Higaki and T.~Kobayashi,
Phys.\ Rev.\ D {\bf 68}, 046006 (2003)
[arXiv:hep-th/0304200].

\bibitem{Aldazabal:2000sa}
G.~Aldazabal, L.~E.~Ibanez, F.~Quevedo and A.~M.~Uranga,
JHEP {\bf 0008}, 002 (2000)
[arXiv:hep-th/0005067]; 
%
J.~F.~G.~Cascales, M.~P.~Garcia del Moral, F.~Quevedo and A.~M.~Uranga,
JHEP {\bf 0402}, 031 (2004)
[arXiv:hep-th/0312051].








\bibitem{Becker:2002nn}
K.~Becker, M.~Becker, M.~Haack and J.~Louis,
JHEP {\bf 0206}, 060 (2002)
[arXiv:hep-th/0204254]; 
%
J.~P.~Conlon, F.~Quevedo and K.~Suruliz,
JHEP {\bf 0508}, 007 (2005)
[arXiv:hep-th/0505076]; 
%
G.~von Gersdorff and A.~Hebecker,
Phys.\ Lett.\ B {\bf 624}, 270 (2005)
[arXiv:hep-th/0507131];
%
M.~Berg, M.~Haack and B.~Kors,
arXiv:hep-th/0508043; 
%
M.~Berg, M.~Haack and B.~Kors,
arXiv:hep-th/0508171; 
%
T.~Higaki, Y.~Kawamura, T.~Kobayashi and H.~Nakano,
Phys.\ Lett.\ B {\bf 582}, 257 (2004)
[arXiv:hep-ph/0311315]; 
%
T.~Higaki, T.~Kobayashi and O.~Seto,
JHEP {\bf 0407}, 035 (2004)
[arXiv:hep-ph/0402200].

\bibitem{Loaiza-Brito:2005fa}
O.~Loaiza-Brito, J.~Martin, H.~P.~Nilles and M.~Ratz,
arXiv:hep-th/0509158.





\bibitem{Krasnikov:1987jj}
See for example,  N.~V.~Krasnikov,
  %
  Phys.\ Lett.\ B {\bf 193}, 37 (1987); 
%
  T.~R.~Taylor,
  %
  Phys.\ Lett.\ B {\bf 252}, 59 (1990); 
%
  J.~A.~Casas, Z.~Lalak, C.~Munoz and G.~G.~Ross,
  %
  Nucl.\ Phys.\ B {\bf 347}, 243 (1990); 
%
  B.~de Carlos, J.~A.~Casas and C.~Munoz,
  %
  Nucl.\ Phys.\ B {\bf 399}, 623 (1993)
  [hep-th/9204012]; 
%
  M.~Dine and Y.~Shirman,
  %
  Phys.\ Rev.\ D {\bf 63}, 046005 (2001)
  [hep-th/9906246];
%
  T.~Banks and M.~Dine,
  %
  Phys.\ Rev.\ D {\bf 50}, 7454 (1994)
  [hep-th/9406132]; 
%
  P.~Binetruy, M.~K.~Gaillard and Y.~Y.~Wu,
  %
  Nucl.\ Phys.\ B {\bf 481}, 109 (1996)
  [hep-th/9605170]; 
%
  J.~A.~Casas,
  %
  Phys.\ Lett.\ B {\bf 384}, 103 (1996)
  [hep-th/9605180]; 
%
  K.~Choi, H.~B.~Kim and H.~D.~Kim,
  Mod.\ Phys.\ Lett.\ A {\bf 14}, 125 (1999)
  [hep-th/9808122];
%
J.~J.~Blanco-Pillado {\it et al.},
JHEP {\bf 0411}, 063 (2004)
[arXiv:hep-th/0406230].




\bibitem{Becker:2004gw}
M.~Becker, G.~Curio and A.~Krause,
Nucl.\ Phys.\ B {\bf 693}, 223 (2004)
[arXiv:hep-th/0403027].

\bibitem{Curio:2001qi}
G.~Curio and A.~Krause,
Nucl.\ Phys.\ B {\bf 643}, 131 (2002)
[arXiv:hep-th/0108220].



\bibitem{Brustein:1992nk}
R.~Brustein and P.~J.~Steinhardt,
Phys.\ Lett.\ B {\bf 302}, 196 (1993)
[arXiv:hep-th/9212049].

\bibitem{kapusta}
J.I.~Kapusta, ``Finite Temperature Field Theory'', 
Cambridge, 1989.


\bibitem{Buchmuller:2004xr}
W.~Buchmuller, K.~Hamaguchi, O.~Lebedev and M.~Ratz,
Nucl.\ Phys.\ B {\bf 699}, 292 (2004)
[arXiv:hep-th/0404168]; 
JCAP {\bf 0501}, 004 (2005)
[arXiv:hep-th/0411109].







\end{thebibliography}
\end{document}